\documentclass[prd,twocolumn,superscriptaddress,altaffilletter,showpacs,nofootinbib]{revtex4}
\usepackage[dvips]{graphicx}
\usepackage{amsmath}

\newcommand{\be}{\begin{equation}}
\newcommand{\ee}{\end{equation}}
\newcommand{\bea}{\begin{eqnarray}}
\newcommand{\eea}{\end{eqnarray}}
\newcommand{\der}{\partial}
\newcommand{\vphi}{\varphi}

\begin{document}

\title{Phase Space Dynamics of Non-Gravitational Interactions between Dark Matter and Dark Energy: The Case of Ghost Dark Energy}

\author{Ricardo Garc\'{\i}a-Salcedo}\email{rigarcias@ipn.mx}\affiliation{Centro de Investigacion en Ciencia Aplicada y Tecnologia Avanzada - Legaria del IPN, M\'exico D.F., M\'exico.}

\author{Tame Gonzalez}\email{tamegc72@gmail.com}\affiliation{Departamento de Ingenier\'ia Civil. Divisi\'on de Ingenier\'ia. Campus Guanajuato. Universidad de Guanajuato.}

\author{Israel Quiros}\email{iquiros6403@gmail.com}\affiliation{Departamento de Matem\'aticas, Centro Universitario de Ciencias Ex\'actas e Ingenier\'ias, Corregidora 500 S.R., Universidad de Guadalajara, 44420 Guadalajara, Jalisco, M\'exico.}

\date{\today}

\begin{abstract}
We study the phase space asymptotics of the so called Veneziano ghost dark energy models. Models where the ghost field's energy density: i) $\rho_{ghost}\propto H$, and ii) $\rho_{ghost}\propto H+H^2$, are investigated. Both, cases with and without additional non-gravitational interaction between cold dark matter and ghost dark energy, are subject to scrutiny. We pay special attention to the choice of phase space variables leading to bounded and compact phase space so that no critical point of physical interest is missing. A rich asymptotic structure is revealed: depending on the kind of non-minimal coupling critical points associated with radiation dominance, matter dominance, cold dark matter/ghost dark energy scaling, and ghost dark energy dominance, are found. Past and future attractors, as well as saddle equilibrium points, are identified in the corresponding phase spaces.
\end{abstract}

\pacs{02.30.Hq, 02.40.Vh, 02.60.Lj, 04.20.-q, 04.50.Kd, 05.45.-a, 05.70.Jk, 95.35.+d, 95.36.+x, 98.80.-k, 98.80.Bp, 98.80.Cq, 98.80.Jk}
\maketitle

\section{Introduction}

The origin and nature of the present stage of accelerated expansion of the universe persists as one of the unexplained mysteries in theoretical physics. Assuming Einstein's general relativity (GR) as the classical theory of the gravitational interactions leads to a large gallery of models which explain the current inflationary stage of the cosmic evolution within appropriate data accuracy (see \cite{de-models} for well-known reviews on the subject). The so called 'ghost' dark energy (GDE) models \cite{urban,ariel,ohta,chinos,chinos',also,instability,alberto-rozas,other,gde-thermodynamics,iterms-also,global-behaviour} have joined the last this vast gallery.\footnote{From the cosmological point of view ghost dark energy is a particular case of dark energy models with generalized equations of state \cite{odintsov}.} According to this proposal the cosmological constant arises from the contribution of the so called Veneziano ghost fields, which are required for the resolution of the U(1) problem in the low-energy effective theory of quantum chromo-dynamics (QCD) \cite{low-e-qcd}. Although the Veneziano ghosts are unphysical and make no contribution in flat Minkowski spacetime, if curved or time-dependent spacetime backgrounds are considered, the cancellation of their contribution to the vacuum energy leave a small energy density $\rho\sim\Lambda^3_{QCD}H$, where $H$ is the Hubble parameter and $\Lambda_{QCD}\sim 100 MeV$ is the QCD mass scale \cite{urban,ariel,ohta} (see also \cite{gde-motivation,gde-motivation'}). If assume $H\sim 10^{-33} eV$ the vacuum energy contribution from the ghost fields gives the right magnitude $\sim(3\times10^{-3} eV)^4$ \cite{ohta}. This lucky circumstance is what have stimulated the interest of cosmologists in these models of dark energy. Their most attractive feature resides in the fact that nothing behind the standard model (SM) of particles and general relativity is required to explain the origin of the dark energy.

The basic ideas of the model were put forth in \cite{urban,ariel} by means of a simplified model. In reference \cite{ohta} a more realistic model was considered and it was shown that indeed QCD ghosts produce dark energy in the right amount. The cosmological dynamics of a simple GDE model where the energy density of the DE is proportional to the Hubble parameter $\rho_{gde}=\alpha H$ ($\alpha\sim\Lambda^3_{QCD}$) has been studied in Ref.\cite{chinos}. In this model the universe has a de Sitter phase at late time and begins to accelerate at redshift around $z_*\sim 0.6$. The authors fit the model and give the constraints on the model's parameters with current observational data including SnIa, BAO, CMB, BBN and Hubble parameter data. The impact of additional non-gravitational interaction between GDE and cold dark matter (CDM) on the cosmic dynamics was also investigated in \cite{chinos}. It was argued therein that, since the squared sound speed of GDE $c^2_s$ is negative, this may cause an instability. The issue was investigated in detail in Ref.\cite{instability}. It was found that, due to non-positivity of the squared sound speed, both non-interacting and interacting GDE models are classically unstable against perturbations in flat and non-flat Friedmann-Robertson-Walker (FRW) backgrounds. The seeming instability problem of GDE models raised by negativity of $c^2_s$ is overcame by means of a simple yet solid argument \cite{alberto-rozas}: unlike other ghost dark energy models, where the ghost field becomes a real propagating physical degree of freedom subject to stringent constraints (see Ref.\cite{piazza}), the Veneziano ghost is not a new physical propagating degree of freedom and the corresponding GDE model does not violate unitarity, causality, or gauge invariance as clearly stated in \cite{ariel}. 

Models of GDE where $\rho_{gde}=\alpha H+\beta H^2$ (the term $\propto H^2$ is inspired by a decaying cosmological constant \cite{decaying-lambda,maggiore}) were explored in \cite{chinos'} (see also \cite{also}). Other aspects of GDE models such as: i) equivalence with kinetic k-essence \cite{alberto-rozas}, ii) connection with Brans-Dicke, and $f(R)$ theories, and tachyons \cite{other}, and iii) thermodynamical issues \cite{gde-thermodynamics}, have been also investigated. In the latter case the GDE energy density $\rho_{gde}$ is shown to be related with the radius of the trapping horizon $\tilde r_T$ \cite{gde-thermodynamics}: $$\rho_{gde}=\frac{\alpha(1-\epsilon)}{\tilde r_T}=\alpha(1-\epsilon)\sqrt{H^2+\frac{k}{a^2}},$$ where $\epsilon\equiv\dot{\tilde r}_T/2H\tilde r_T$. If neglect spatial curvature, as we do in this paper, the trapping horizon is coincident with the Hubble horizon $\tilde r_T=1/H$, and $\rho_{gde}=\alpha(1-\epsilon)H$. Here we also neglect the contribution $\propto\epsilon$ as it is done in most works on GDE \cite{urban,ariel,ohta,chinos,also,instability,alberto-rozas,other,gde-thermodynamics,iterms-also}. This particular case, strictly speaking, corresponds to dark energy dominance so that, in general, the contribution of $\epsilon$ to the dynamics has to be taken under consideration, as it is done in \cite{global-behaviour}. In that reference the global dynamical behavior of the universe accelerated by the QCD GDE, where $\rho_{gde}=\alpha(1-\epsilon)\sqrt{H^2+k/a^2}$, was investigated by using the dynamical systems tools. Depending on the values of free parameters only two critical points of physical relevance were found.\footnote{Dependence of the phase space structure on the free parameters signals bifurcation which is a very interesting finding in Ref.\cite{global-behaviour}.} Due to the choice of phase space variables in \cite{global-behaviour}: $\mu=\Omega_{cdm}/\Omega_{gde}$ ($\Omega_i\equiv\rho_i/3H^2$) and $\epsilon$, both points are correlated with CDM/GDE-scaling behavior since in both cases $\mu=\mu_c\neq 0$.   

The dynamical systems tools supply powerful means to extract very useful information out of cosmological models without worrying about exact solutions. There is an isomorphism between particular solutions of the cosmological field equations and points of the equivalent phase space. Critical (also equilibrium or fixed) points in the phase space can be correlated with particular (classes of) solutions of the cosmological equations. If the critical point were either a future or a past attractor, then, the corresponding cosmological solution were a generic or preferred solution of the cosmological equations since, quite independent on the initial conditions, orbits in the phase space start at a past attractor and end up at a future attractor whenever these attractors exist. However, no matter how useful these tools are, one have to take with care the results of the linear dynamical systems study and never should overestimate their reach. It is central for the linear study, in particular, that the equivalent phase space be bounded and compact whenever it is possible. Only in this case all of the relevant equilibrium points are 'visible'. Otherwise several critical points may be missing and one will need of additional phase space variables to cover the region not covered by the original variables (see section III below). This is precisely what we think happens in the study of reference \cite{global-behaviour}. Actually, due to the definition of the variable $\mu$ (see above), the phase space in the mentioned reference is the unbounded plane region $\{(\mu,\epsilon)|0\leq\mu<\infty,\,\epsilon<1\}$. Hence, several critical points of physical relevance, in particular those associated with cosmological solutions which do not represent scaling behavior, may be missing in \cite{global-behaviour}.

The aim of the present paper is the detailed investigation of the phase space dynamics of GDE cosmological models with/without additional non-gravitational interaction with CDM, by means of the dynamical systems tools. We shall consider GDE models where: i) $\rho_{gde}=\alpha H$ \cite{ohta,chinos,instability,alberto-rozas,other,gde-thermodynamics,iterms-also,global-behaviour}, and ii) $\rho_{gde}=\alpha H+\beta H^2$ \cite{chinos',also}. Due to the amount of results to discuss, the most realistic case when $\rho_{gde}=\alpha(1-\epsilon)H$ will be left for a sequel to this work. Our goal here will be to uncover the main features of the asymptotics of these cosmological models within the equivalent phase space so that we can correlate equilibrium points (past/future attractors and saddle points) with their generic cosmological solutions. We pay special attention to the choice of phase space variables which lead to bounded and compact physically meaningful phase space, so that no critical point of cosmological interest is missing. The arising asymptotic structure of the GDE models is richer than the one found in Ref.\cite{global-behaviour}. In the interacting case, depending on the kind of assumed interaction, besides the already mentioned CDM/GDE scaling equilibrium points, also critical points associated with radiation dominance, matter dominance, and GDE dominance are found.

We start by illustrating the method through investigating the dynamics of a well-known physical system in section II: a quintessence model with an exponential self-interaction potential. In this context we shall look at the role non-gravitational interactions play in the cosmic dynamics by comparing with well-known results for the non-interacting case \cite{wands}. Non-gravitational interactions have proven useful to overcome the coincidence problem in the context of GR models of dark energy \cite{amendola,interaction,nm-coupling}. However, the success of the interaction in solving the problem depends on the assumed kind of interaction. Here we shall consider specific kinds of interaction commonly studied in the bibliography \cite{iterms1,iterms2,iterms-hde}. In section III we give brief tips on (linear) phase space analysis which will be useful latter on when we explore the phase space dynamics of GDE models. Then we apply the methodology used in section II to explore the ghost dark energy models $\rho_{gde}\propto H$ in section IV, and $\rho_{gde}\propto H+H^2$ in section V. The results of our study and their implications will be discussed in section VI, while brief conclusions will be supplied in the final section VII. Through the paper we use natural units ($8\pi G\equiv8\pi/m_{PL}^2\equiv c\equiv1$).

\section{Interacting Quintessence}\label{quintessence}

In this section we shall discuss in a more or less pedagogical way, how the quintessential cosmological dynamics driven by an exponential self-interaction potential looks like in the equivalent phase space. Although the phase space dynamics of non-interacting quintessence fields has been investigated in detail \cite{wands}, the phase space description of the interacting case is less known (see, however \cite{amendola,interaction,nm-coupling}). In the last instance, the discussion in the present section will serve as an illustration of the method and terminology we shall apply to the investigation of the ghost dark energy case in the next sections. Throughout the paper we shall consider Friedmann-Robertson-Walker (FRW) spacetimes with flat spatial sections, $$ds^2=-dt^2+a^2(t)\delta_{ik}dx^idx^k,\;\;\;i,k=1,2,3.$$ 

For simplicity we shall assume that the cosmic dynamics is fueled by a mixture of: i) pressureless CDM with energy density $\rho_m$, and, ii) dark energy. In this section the dark energy component is in the form of a quintessence field $\vphi$, with energy density $\rho_\vphi$, and parametric pressure $p_\vphi$ given by: 

$$\rho_\vphi=\dot\vphi^2/2+V(\vphi),\;p_\vphi=\dot\vphi^2/2-V(\vphi),$$ respectively. In these equations $V=V(\vphi)$ is the self-interacting potential for quintessence. We also assume that the CDM and the quintessence field interact through an additional force of non-gravitational origin.

The quintessence field equations are the following:

\bea &&3H^2=\rho_m+\rho_\vphi,\nonumber\\
&&\dot\rho_\vphi+3H(\omega_\vphi+1)\rho_\vphi=-Q,\nonumber\\
&&\dot\rho_m+3H\rho_m=Q,\label{q-feqs}\eea where $Q$ stands for the non-gravitational interaction term, and $\omega_\vphi$ -- the quintessence equation of state (EOS) parameter:

\be \omega_\vphi\equiv\frac{p_\vphi}{\rho_\vphi}=\frac{\dot\vphi^2-2V}{\dot\vphi^2+2V}.\label{q-eos}\ee

Our goal will be to study the dynamics dictated by equations (\ref{q-feqs}) in the equivalent phase space. In order to obtain an autonomous system of ordinary differential equations (ODE) out of (\ref{q-feqs}) we introduce the following dimensionless phase space variables \cite{wands}:

\be x\equiv\frac{\dot\vphi}{\sqrt 6 H},\;y\equiv\frac{\sqrt V}{\sqrt 3 H},\label{q-ps-var}\ee which amount to the (square of the) dimensionless kinetic and potential energy density of the quintessence field respectively. In terms of these variables the Friedmann constraint (first equation in (\ref{q-feqs})) can be written as

\be \Omega_m\equiv\frac{\rho_m}{3H^2}=1-x^2-y^2,\label{friedmann-c}\ee while $$\frac{H'}{H}=-\frac{3}{2}(1+x^2-y^2),$$ where the tilde accounts for derivative with respect to the parameter, $\tau\equiv\int da/a\;\Rightarrow\;d\tau=d\ln a=H dt$. 

Since $0\leq\Omega_m\leq 1$, then, it follows that, $0\leq x^2+y^2\leq 1$. We shall be concerned here with cosmological expansion ($H\geq 0$), so that, only non-negative values of the variable $y$ ($y\geq 0$) will be considered.

Other physical parameters of interest are:

\bea &&\Omega_\vphi\equiv\frac{\rho_\vphi}{3H^2}=x^2+y^2,\;\omega_\vphi=\frac{x^2-y^2}{x^2+y^2},\nonumber\\
&&q=-\left(1+\frac{H'}{H}\right)=\frac{1}{2}+\frac{3}{2}(x^2-y^2).\label{p-parameters}\eea 

The following autonomous system of ODE is obtained out of the original set of cosmological equations:

\bea &&x'=-\frac{3}{2}x(1+y^2-x^2)-\sqrt\frac{3}{2}\frac{V_{,\vphi}}{V}y^2-\frac{Q}{6 H^3 x},\nonumber\\
&&y'=\frac{3}{2}y\left(1+x^2-y^2+\sqrt\frac{2}{3}\frac{V_{,\vphi}}{V}x\right).\label{q-asode}\eea

Depending on the assumed kind of interaction $Q$, and of the self-interacting potential $V$, one might need to introduce one or two additional phase space variables. For convenience the interaction term can be written as $Q=H\delta$, where $\delta$ is an energy density ascribed to the interaction. If $\delta>0$, the transfer of energy is from the DE to the CDM component and vice versa.

Here we shall concentrate on the exponential potential, $V=V_0\;e^{-\mu\vphi}$ ($V_0$ and $\mu$ are constants), which is one of the most common potentials found in the applications to scalar field models of dark energy. In this case $V_{,\vphi}/V=-\mu$. We shall assume different forms of the interaction which are frequently found in the bibliography on the subject \cite{iterms1,iterms2,iterms-also}.

\subsection{$\delta=\delta_0=const.$ \cite{iterms1}.}\label{q-1}

Here we shall assume positive $\delta_0>0$, so that the transfer of energy is from the quintessence field into the CDM. In this case it is convenient to introduce an additional variable,

\be z\equiv\frac{3H^2}{3H^2+\delta_0},\;0\leq z\leq 1,\label{z}\ee to the variables in (\ref{q-ps-var}). Notice that the value $z=1$ corresponds to either, i) no interaction ($\delta_0=0$), or ii) to the initial singular state ($H^2\rightarrow\infty$). Meanwhile $z=0$ $\Rightarrow$ $H=0$, which means in turn that $Q=H\delta_0=0$, i. e., the interaction vanishes. We have $$Q=3H^3\left(\frac{1-z}{z}\right).$$ Accordingly, the autonomous system of ODE one obtains can be written in the following form:

\bea &&x'=\sqrt\frac{3}{2}\mu\;y^2-\frac{3}{2}x(1+y^2-x^2)+\frac{z-1}{2xz},\nonumber\\
&&y'=\frac{3}{2}y\left(1+x^2-y^2-\sqrt\frac{2}{3}\mu\;x\right),\nonumber\\
&&z'=3z(z-1)(1+x^2-y^2).\label{q-0-asode}\eea 

The 3D phase space where to look for critical points of (\ref{q-0-asode}) can be defined as the following compact half-cylinder with unit height and unit radius:

\be \Psi_{3D}=\{(x,y,z)|0\leq x^2+y^2\leq 1, y\geq 0, 0\leq z\leq 1\}.\label{q-3d-ps}\ee

We found five critical points $P_{c_i}:(x_{c_i},y_{c_i},z_{c_i})\in\Psi_{3D}$, which correspond to the five critical points found in \cite{wands} for the non-interacting case. The stability properties of these points, however, are modified by the non-gravitational interaction between the dark matter and the quintessence field. The main properties of the equilibrium points are listed below.

\begin{enumerate}

\item Stiff matter-dominated solution(s), $$P^\pm_{st}:(\pm 1,0,1),\;\Omega_\vphi=1.$$ The quintessence field behaves like stiff matter, $\omega_\vphi=1$, and the expansion is highly decelerated, $q=2$. The eigenvalues of the Jacobian (also linearization) matrix, $$\begin{pmatrix} \frac{\der x'}{\der x} & \frac{\der x'}{\der y} & \frac{\der x'}{\der z} \\ \frac{\der y'}{\der x} & \frac{\der y'}{\der y} & \frac{\der y'}{\der z} \\ \frac{\der z'}{\der x} & \frac{\der z'}{\der y} & \frac{\der z'}{\der z} \end{pmatrix}_{P^\pm_{st}}=\begin{pmatrix} 3 & 0 & \frac{1}{2} \\ 0 & 3\mp\sqrt\frac{3}{2}\mu & 0 \\ 0 & 0 & 6 \end{pmatrix},$$ are: $\lambda_1=3,\;\lambda_2=3\mp\sqrt{3/2}\mu,\;\lambda_3=6$. This means that, if $\mu<\sqrt 6$, $P^+_{st}$ is an unstable (source) equilibrium point, also called ''past attractor'' because the orbits in the phase space approach to this point into the past. In case if $\mu>\sqrt 6$, $P^+_{st}$ is a saddle critical point instead, which corresponds to absence of non-gravitational interactions ($\delta_0=0$). Meanwhile, $P^-_{st}$, is always a past attractor. Since, $z=1$, this solution can be associated either with the absence of additional non-gravitational interaction ($\delta_0=0$), or, with the initial singular state with diverging, $H\rightarrow\infty$. 

\item CDM-dominated point, $$P_m:(0,0,1),\;\Omega_m=1.$$ This equilibrium point corresponds to decelerated expansion, $q=1/2$, and the quintessence EOS parameter, $\omega_\vphi$, is undefined. Since the eigenvalues of the corresponding linearization matrix, $\lambda_1=3/2,\;\lambda_2=-3/2,\;\lambda_3=3$, are of opposite sign, this means that this is a saddle critical point. A saddle in the phase space is ''meta-stable'' in the sense that it is unstable in some directions ($x$ and $z$-directions in the present case), while being stable in other directions in phase space ($y$-direction in this case). Orbits in $\Psi_{3D}$ which approach to the saddle point will evolve for some $\tau$-time in its neighborhood, until they are eventually repelled from it. This would mean that the saddle point corresponds to a transient stage of the cosmic evolution. In this case the matter-dominated phase is transient as it should be to account for the observed amount of cosmic structure. 

\item CDM/quintessence-scaling solution, $$P_{m-sc}:\left(\frac{\sqrt{3/2}}{\mu},\frac{\sqrt{3/2}}{\mu},1\right)\;\Rightarrow\;\frac{\Omega_m}{\Omega_\vphi}=\frac{1-3/\mu^2}{3/\mu^2}.$$ This equilibrium point exists whenever $\mu^2\geq 3$. It is associated with decelerating expansion ($q=1/2$). The quintessence field mimics the behavior of pressureless matter ($\omega_\vphi=0$). The eigenvalues of the corresponding Jacobian matrix are: $$\lambda_{1,2}=\frac{3}{4}\left(-1\pm\sqrt{\frac{24}{\mu^2}-7}\right),\;\lambda_3=3.$$ Hence, this is also a saddle point in $\Psi_{3D}$, so that the scaling solution describes a transient stage of the cosmic evolution. For $\mu^2>24/7$, this is a spiral critical point.

\item Quintessence-dominated solution, $$P_q:\left(\mu/\sqrt 6,\sqrt{1-\mu^2/6},1\right)\;\Rightarrow\;\Omega_\vphi=1,$$ also known as potential/kinetic energy-scaling solution since the potential and the kinetic energies of the quintessence field scale as: $$\frac{V}{\dot\vphi^2/2}=\frac{y^2}{x^2}=\frac{1-\mu^2/6}{\mu^2/6}.$$ This equilibrium point exists if $\mu^2<6$. In this case, $$q=-1+\mu^2/2,\;\omega_\vphi=-1+\mu^2/3,$$ so that, the corresponding cosmological solution is accelerated whenever $\mu^2<2$. Otherwise it is decelerating. The eigenvalues of the Jacobian matrix corresponding to this point are: $\lambda_1=-3+\mu^2/2,\;\lambda_2=-3+\mu^2,\;\lambda_3=\mu^2$, so that, it is also a saddle critical point in $\Psi_{3D}$.

\end{enumerate}

Comparing our results with the results of reference \cite{wands}, shows that, while there are not found new equilibrium points in the phase space in respect with the non-interacting case, the stability of the critical points $P_{m-sc}$, and $P_q$, has been indeed modified by the additional (non-gravitational) interaction between CDM and quintessence components. Actually, while in the absence of interaction, depending on the value of the parameter $\mu^2$, the critical point $P_{m-sc}$ can be, either a stable spiral ($\mu^2>24/7$), or a stable node ($3<\mu^2<24/7$), in the presence of constant interaction, $Q=H\delta_0$, $P_{m-sc}$ is always a saddle critical point. Similarly, in the absence of interaction the point $P_q$ is a stable node whenever $\mu^2<3$ and a saddle if $3<\mu^2<6$. If we set-on the constant additional interaction, $Q=H\delta_0$, of non-gravitational origin, due to the fact that one of the eigenvalues, $\lambda_3$, is always positive, while the other, $\lambda_1$, is always negative, $P_q$ is always a saddle point.

Note that all of the critical points of the autonomous system of ODE (\ref{q-0-asode}), lay on top ($z=1$) of the half-cylinder $\Psi_{3D}$ (\ref{q-3d-ps}): $(x,y,1)\in\Psi_{3D}$. I. e., there are not critical points associated with the additional non-gravitational interaction, $Q=H\delta_0$. 

In the next subsections we shall look for modifications to the phase space dynamics of the present quintessence model which are induced by setting on an additional non-gravitational interaction between CDM and quintessence, in two generic ways found in the bibliography \cite{iterms1,iterms2,iterms-also}, where $\delta\neq const$.

\subsection{$\delta=3b^2\rho_m$ \cite{iterms1}.}\label{q-2}

Here $b^2$ is a non-negative constant. In this case the $x$ and $y$ variables in Eq.(\ref{q-ps-var}) are enough and, besides: $$\frac{Q}{6H^3}=\frac{3b^2}{2}(1-x^2-y^2).$$ Hence, the autonomous system of ODE (\ref{q-asode}) can be written as:

\bea &&x'=-\frac{3}{2}x(1+y^2-x^2)+\sqrt\frac{3}{2}\mu\;y^2\nonumber\\
&&\;\;\;\;\;\;\;\;\;\;\;\;\;\;\;\;\;\;\;\;\;\;\;\;\;\;-\frac{3b^2}{2x}(1-x^2-y^2),\nonumber\\
&&y'=\frac{3}{2}y\left(1+x^2-y^2-\sqrt\frac{2}{3}\mu\;x\right).\label{q-2-asode}\eea 

We shall seek for critical points of (\ref{q-2-asode}) within the unit half-disc:

\be \Psi_{2D}=\{(x,y)|0\leq x^2+y^2\leq 1, y\geq 1\}.\label{q-2d-ps}\ee

Four critical points $P_{c_i}:(x_{c_i},y_{c_i})\in\Psi_{2D}$ are found.

\begin{enumerate}

\item Stiff matter solution(s): $$P^\pm_{st}:(\pm 1,0),\;\omega_\vphi=1,\;q=2,$$ which is dominated by the quintessence field since, $\Omega_\vphi=1$. The eigenvalues of the Jacobian matrix for $P^\pm_{st}$ are, $\lambda_1=3(1+b^2)$, and $\lambda^\pm_2=3\mp\sqrt{3/2}\mu$, respectively. Stability of these points is not affected in a significant way (with respect to the results of \cite{wands}) by the non-gravitational interaction, $Q=3b^2H\rho_m$. 

\item CDM/quintessence scaling solution: $$P_{m-sc}:\left(\sqrt\frac{3}{2}\frac{(1-b^2)}{\mu},\sqrt\frac{3}{2}\frac{(1-b^2)}{\mu}\sqrt{1+\frac{2b^2\mu^2}{3(1-b^2)^2}}\;\right).$$ The point exists whenever, $b^2\leq 1$, $\mu^2\geq 3(1-b^2)$. In this case CDM and quintessence scale as, $$\frac{\Omega_m}{\Omega_\vphi}=\frac{1-(b^2+3(1-b^2)^2/\mu^2)}{b^2+3(1-b^2)^2/\mu^2}.$$ The quintessence EOS parameter, and the deceleration parameter are given by: $$\omega_\vphi=-\frac{\mu^2 b^2}{\mu^2 b^2+3(1-b^2)^2},\;q=\frac{1}{2}-\frac{3b^2}{2},$$ respectively. Note that, although both components of the cosmic fluid scale in a constant fraction, the quintessence does not exactly mimic the pressureless CDM, since, $\omega_\vphi=0$, only in the absence of interaction ($b^2=0$). The eigenvalues of the linearization matrix are: $$\lambda_{1,2}=-\frac{\alpha\pm\sqrt\beta}{4(1-b^2)\mu},$$ where we have defined, $\alpha:=-2\mu^3b^2+3(1-b^2)(1+3b^2)\mu$, and $\beta:=4\mu^6b^4-12b^2(1-b^2)(5-b^2)\mu^4-9(1-5b^2)(7-3b^2)\mu^2+216(1-b^2)^5$. Here we will not analyze in details the stability of this point but it is intuitive to see that the interaction does not modify in an appreciable way the results of \cite{wands} but for the fact that, if $b^2>1/3$, then $q<0$, and the CDM/quintessence scaling solution may be correlated with accelerated expansion \cite{amendola}.

\item Quintessence-dominated point ($\Omega_\vphi=1$), $$P_q:\left(\mu/\sqrt 6,\sqrt{1-\mu^2/6}\right),\;\omega_\vphi=-1+\mu^2/3,$$ which exists if $\mu^2<6$. The deceleration parameter, $q=-1+\mu^2/2$. The eigenvalues of the linearization matrix for this critical point are: $\lambda_1=-3+\mu^2/2$, $\lambda_2=-3(1-b^2)+\mu^2$. As seen, in this case, like in the former two ones, neither the stability, nor the existence conditions for the quintessence-dominated critical point are modified in any sensible way by the additional (non-gravitational) interaction, $Q=3b^2H\rho_m$.

\end{enumerate}

We can summarize this subsection by noting that, although the stability and existence conditions for the critical points are not appreciably modified by the interaction term, $Q=3b^2H\rho_m$, the topology of the phase space itself is indeed modified since one fixed point is missing with respect to the non-interacting case. Actually, it is easy to note that the CDM-dominated solution of the cosmological equations is not a fixed point of the equivalent autonomous system of ODE, as it is the case in the absence of interaction \cite{wands}. The CDM/scaling point $P_{m-sc}$ also differs from the corresponding one in the non-interacting case, since, i) the expansion may be accelerating as long as $b^2>1/3$, and ii) the quintessence EoS parameter $\omega_\vphi\neq 0$, i. e., it does not track the CDM equation of state. This means that a mix of two different fluids: i) dust CDM, and ii) a scalar fluid with a different equation of state, co-exist during this stage of the cosmic evolution, in such a way that the ratio of their energy densities is a constant. 

A matter(CDM)-dominated point, or, in its absence, a matter/quintessence scaling point with $\omega_\vphi=0$, is necessary to account for the observed amount of cosmic structure. Hence, the addition of an interaction term of the form, $Q=3b^2H\rho_m$, to the quintessence field equations, transforms a suitable model of dark energy (as it is, in particular, the quintessence model driven by an exponential potential) into a model which is ruled out by observational data on structure formation.

\subsection{$\delta=3b^2(\rho_m+\rho_\vphi)$ \cite{iterms2}.}\label{q-3}

Finally, to end up this 'illustrative' section, we shall briefly consider an interaction term which generalizes the former one, and has been studied in detail in several cosmological contexts \cite{iterms2}. 

As before the phase space coincides with the unit half-disc, $\Psi_{2D}$, of equation (\ref{q-2d-ps}). Since, $Q/6H^3=3b^2/2$, the autonomous system of ODE (\ref{q-asode}) simplifies to:

\bea &&x'=-\frac{3}{2}x(1+y^2-x^2)+\sqrt\frac{3}{2}\mu\;y^2-\frac{3b^2}{2x},\nonumber\\
&&y'=\frac{3}{2}y\left(1+x^2-y^2-\sqrt\frac{2}{3}\mu\;x\right).\label{q-3-asode}\eea 

Without going into details, one founds that, of the five equilibrium points found in the non-interacting case \cite{wands}, only three critical points remain: i,ii) the stiff matter solution(s), $$P^\pm_{st}:\left(\pm\sqrt\frac{1+\sqrt{1+4b^2}}{2},0\right),$$ and, iii) the CDM/quintessence scaling point, $$P_{m-sc}:(x_c,y_c),\;\frac{\Omega_m}{\Omega_\vphi}=\frac{1-x_c^2-y_c^2}{x_c^2+y^2_c},\;\omega_\vphi=\frac{x^2_c-y^2_c}{x^2_c+y^2_c},$$ where, $x_c$, and, $y_c$, are the positive roots of the system of algebraic equations:

\bea &&1+x^2-y^2-\sqrt\frac{2}{3}\mu x=0,\nonumber\\
&&x^2(1+y^2-x^2)-\sqrt\frac{2}{3}\mu x y^2+b^2=0.\nonumber\eea 

Hence, if consider an interaction term of the form $Q=3b^2H(\rho_m+\rho_\vphi)$, two important critical points are erased from the phase space by the non-gravitational interaction between CDM and DE: i) the (transient) CDM-dominated solution, which is necessary for the right formation of structure, and, ii) the quintessence-dominated solution which is important to get a stage of accelerated expansion which could be correlated with the present period of the cosmic history. 

In the next sections, on the grounds of a similar dynamical systems analysis, we shall investigate the phase space dynamics of the so-called ghost dark energy models, in order to seek for modifications to the corresponding dynamics originated from consideration of additional non-gravitational interactions. We shall be considering, in particular, the specific forms of the interaction discussed in subsections \ref{q-1}, \ref{q-2}, and \ref{q-3}.

\section{Tips on phase space analysis}

There is an isomorphism between particular solutions of the cosmological field equations and points of the equivalent phase space. Take as an example, the quintessence cosmological model with exponential potential studied in the former section, with additional non-gravitational interaction in the form of (subsection \ref{q-1}): $Q=H\delta_0$. Consider the quintessence-dominated solution, $$P_q:\left(\mu/\sqrt 6,\sqrt{1-\mu^2/6},1\right).$$ In this case, since, $z=1$, then, $$z=\frac{3H^2}{3H^2+\delta_0}=1\;\Rightarrow\;\delta_0=0.$$ Hence, this point will correspond to a particular solution of the quintessence field equations without non-gravitational interaction between the CDM and the DE. Besides, $$x=\frac{\dot\vphi}{\sqrt 6 H}=\frac{\mu}{\sqrt 6}\;\Rightarrow\;H=\frac{\dot\vphi}{\mu}\;\Rightarrow\;a(\vphi)=a_0 e^{\vphi/\mu},$$ and, $y=\sqrt V/\sqrt 3 H=\sqrt{1-\mu^2/6}$ which leads to $\sqrt{2V}/\dot\vphi=\sqrt{6-\mu^2}/\mu$. The latter equation can be integrated in quadratures to give $\vphi(t)=\vphi_0+\ln(t-t_0)^{2/\mu}$, or, in terms of the scale factor: $a(t)=\bar a_0 (t-t_0)^{2/\mu^2},\;\bar a_0\equiv a_0\;e^\vphi_0$. The above solution for the original field variables corresponds to the fixed point $P_q\in\Psi_{3D}$ -- the phase space defined in Eq.(\ref{q-3d-ps}). This critical point exists if $\mu^2<6$ (see subsection \ref{q-1}), i. e., whenever the power $2/\mu^2>1/3$. The corresponding cosmological solution depicts accelerated expansion whenever, $\mu^2<2$, i. e., if the power, $2/\mu^2>1$. $P_q$ is a saddle equilibrium point. This means that there exists a non-empty set of initial conditions, which picks a congruence of neighboring orbits in $\Psi_{3D}$, which approach to the quintessence-dominated point $P_q$, then evolve for some $\tau$-time in the neighborhood of this point, until they are finally repelled from it. Going back to the standard cosmological ''language'' this would mean that the quintessence-dominated phase, depicted in spacetime by the above solution, can be, at most, a transient stage of the cosmic evolution. I. e., this cosmological solution can not decide the fate of the cosmic evolution.

Studying the phase space associated with a given cosmological model is, some times, more useful than looking for exact (and/or numeric) solutions of the cosmological field equations. Actually, while a given exact solution is necessary to evolve the model from given initial data to get concrete predictions to confront which with observations, fixed points in the (equivalent) phase space tell us, which solutions in the (perhaps infinite) set of possible analytic solutions of the field equations, are generic, or ''preferred'' by the model. Generic or preferred solutions -- corresponding to fixed points in the phase space -- are structurally stable, while the remaining possible solutions are not stable in general. In this last sentence, under ''structural stability'', we understand that the given critical points are reached, either into the future, or into the past (in terms of the time-ordering phase space parameter $\tau$) quite independent of the initial conditions chosen, i. e., the qualitative behavior of the trajectories in the phase space is unaffected by small perturbations. In this regard, a unstable or source critical point can be alternatively understood as a past attractor -- a stable equilibrium point in reversed time direction -- while a stable fixed point is a future attractor.

We want to underline that, even if, according to the mentioned isomorphism, to every (physically meaningful) particular solution of the cosmological field equations of a given model,\footnote{By ''physically meaningful'' we understand the fact that the given solution of the field equations corresponds to a point $(x,y,z)$ which belongs in the phase space $\Psi_{3D}$.} it corresponds a given phase space point, there can be particular solutions of the field equations which can not be associated with fixed points in the phase space, i. e., which correspond to points in the phase space which are not critical. These solutions can be picked up only under very specific initial conditions on the phase space orbits, i. e., these will be structurally unstable solutions. In the subsection \ref{q-2} (also \ref{q-3}), for instance, there are not found fixed points which could be associated with the matter(CDM)-dominated solution. This means that, even if this can be a particular exact solution of the cosmological field equations, it can not describe a matter-dominated stage of the cosmic expansion lasting for enough time as to produce the observed amount of cosmic structure. Under the basis of this analysis, quintessential models with exponential potential, and, with additional non-gravitational interaction of the form $Q=3b^2H\rho_m$ (also, $Q=3b^2H(\rho_m+\rho_\vphi)$) are ruled out.

As we have just illustrated it in a simple quintessence model with an exponential self-interaction potential, a correct choice of the variables of the phase space is a very important point which is, most times, underestimated. One has to be able to choose variables which take values within finite intervals, i. e., one has to be able to have the entire phase space enclosed in a finite region. This guarantees that all of the relevant fixed points will be ''visible'', i. e., none of them will go to infinity. Fortunately, most times, this choice is possible. Consider, for instance, an arbitrary phase space variable $\xi$ taking values in a semi-infinite interval: $0\leq\xi<\infty$. After the following transformation of variable, $$\xi\rightarrow\bar\xi=\frac{1}{1+\xi}\;\;\left(\xi=\frac{1-\bar\xi}{\bar\xi}\right),$$ the above semi-infinite interval is transformed into a finite interval $0<\bar\xi\leq 1$. Using the above prescription one might transform an infinite interval $]-\infty,\infty[$ into a finite one $[-1,0[\;\bigcup\;]0,1]$, by introducing two variables $\bar\xi$ and $\bar{\bar\xi}$ to cover the entire phase space.

In the remainder of this paper we shall replace the quintessence field by the so called QCD ghost dark energy. We shall look for possible modifications to the GDE dynamics by an additional non-gravitational interaction, for the particular choices of the interacting term $Q$ studied in subsections \ref{q-1}, \ref{q-2}, and \ref{q-3}.

\section{Phase Space Dynamics of Ghost Dark Energy}

From now on we shall consider the dark energy component in the form of QCD (Veneziano) ghost \cite{urban,ariel,ohta,chinos,chinos',also,instability,alberto-rozas,other,gde-thermodynamics,iterms-also,global-behaviour}. We start with the simplest case when there is not additional non-gravitational interaction between the GDE and the CDM. For the flat FRW universe filled with GDE, CDM, and radiation, the corresponding Friedmann equation is written as

\be 3H^2=\rho_{gde}+\rho_m+\rho_r,\label{friedman_eq}\ee where $\rho_m$ is the energy density of (pressureless) cold dark matter, $\rho_r$ is the energy density of radiation, and, $\rho_{gde}$ is the GDE energy density assumed to be given by \cite{ohta}, 

\be \rho_{gde}=\alpha H,\label{sim-h}\ee where, since we require non-negative $\rho_{gde}\geq 0$, and since we shall focus in expanding universes only ($H\geq 0$), then, the free parameter $\alpha\geq 0$ will be taken non-negative.

\subsection{Non-interacting Ghost Dark Energy}\label{n-i-gde}

The energy conservation equations for the different components are:

\bea &&\dot\rho_m+3H\rho_m=0,\;\dot\rho_r+4H\rho_r=0,\notag \\
&&\dot\rho_{gde}+3H\rho_{gde}(1+\omega_{gde})=0, \label{conservation_law}\eea where $\omega_{gde}$ is the GDE EOS parameter. The definition for the GDE energy density (\ref{sim-h}), together with Eq.(\ref{conservation_law}), yield to the following relationship: 

\be \frac{\dot\rho_{gde}}{H\rho_{gde}}=\frac{\dot H}{H^2}=-3(1+\omega_{gde}).\label{relation}\ee 

It will be useful to have several quantities written in terms of the dimensionless parameter of GDE energy density $\Omega_{gde}\equiv\rho_{gde}/3H^2$ and of the dimensionless energy density parameter of the radiation component $\Omega_r\equiv\rho_r/3H^2$. In this regard, by taking the time derivative of the Friedmann equation (\ref{friedman_eq}) and, considering equations (\ref{friedman_eq}), and (\ref{conservation_law}), one is left with

\be 3(1+\omega_{gde})=\frac{3-3\Omega_{gde}+\Omega_r}{2-\Omega_{gde}}=-\frac{\dot H}{H^2}.\label{useful}\ee Hence, for the GDE state parameter one obtains

\be \omega_{gde}=-\frac{3-\Omega_r}{3(2-\Omega_{gde})}, \label{stateparameter}\ee while for the deceleration parameter: 
  
\be q=-1+3(1+\omega_{gde})=\frac{1-2\Omega_{gde}+\Omega_r}{2-\Omega_{gde}}.\label{q}\ee

In order to put the cosmological equations in the form of an autonomous system of ODE, we choose appropriate phase space variables: 

\be x\equiv\Omega_{gde},\;\;y\equiv\Omega_r.\label{xy}\ee After this choice the following autonomous system of ODE can be obtained:

\be x'=x\left[\frac{3-3x+y}{2-x}\right],\;y'=-2y\left[\frac{1+x-y}{2-x}\right],\label{ode}\ee where, as before, the tilde denotes derivative with respect to the variable $\tau\equiv\ln a (d\tau=Hdt)$. The phase space relevant to the present study is given by the following triangular (compact) region in ($x,y$)-plane: 

\bea &&\hat\Psi_{2D}=\{(x,y)|0\leq x\leq 1,\;0\leq y\leq 1,\nonumber\\
&&\;\;\;\;\;\;\;\;\;\;\;\;\;\;\;\;\;\;\;\;\;\;\;\;\;\;\;\;\;\;\;\;\;\;\;\;0\leq x+y\leq 1\}.\label{phasespace}\eea In terms of the above variables, the Friedmann equation can be written in the form of the following constraint,

\be \Omega_m=1-x-y,\label{friedmann_constraint}\ee besides, the GDE EOS parameter can be written as

\be \omega_{gde}=\frac{y-3}{3(2-x)}.\label{gde_eos}\ee

Three equilibrium points $P_{c_i}:(x_{c_i},y_{c_i})$ can be found in $\hat\Psi_{2D}$, which correspond to different phases of the cosmic evolution:

\begin{enumerate}

\item Radiation-dominated phase: \be P_r:(0,1),\;\Omega_m=0,\;\Omega_r=1,\;\Omega_{gde}=0.\nonumber\ee This is a decelerating expansion solution ($q=1$). The eigenvalues of the linearization matrix corresponding to this equilibrium point are: $\lambda_1=2$, $\lambda_2=1$, so that it is a unstable critical point (past attractor) in $\hat\Psi_{2D}$. The GDE state parameter is $\omega_{gde}=-1/3$.
  
\item CDM-dominated phase: \be P_m:(0,0),\;\Omega_m=1,\;\Omega_r=0,\;\Omega_{gde}=0.\nonumber\ee This phase of the cosmic evolution is characterized also by decelerated expansion ($q=1/2$). The existence of this solution is necessary for the formation of the observed amount of structure. The eigenvalues of the corresponding linearization matrix are: $\lambda_1=-1$, $\lambda_2=3/2$, so that it is a saddle critical point in $\hat\Psi_{2D}$. For the GDE state parameter we obtain $\omega_{gde}=-1/2$.
  
\item GDE-dominated, de Sitter phase: \be P_{dS}:(1,0),\;\Omega_m=0,\;\Omega_r=0,\;\Omega_{gde}=1.\nonumber\ee This late-time phase corresponds to an inflationary solution ($q=-1$). The eigenvalues of the linearization matrix for $P_{dS}$ are: $\lambda_1=-4$, $\lambda_2=-3$, so that the solution is a future attractor. This equilibrium point mimics cosmological constant behavior since $\omega_{gde}=-1$.
  
\end{enumerate}

According to the global structure of the phase space, orbits in $\hat\Psi_{2D}$ converge towards the radiation-dominated stage into the $\tau$-past, depending on the initial conditions evolve for some time in the neighborhood of the saddle matter-dominated point, until they are repelled from this point to, finally, converge towards the GDE-dominated future (de Sitter attractor). This model correctly describes the fundamental stages of the cosmic evolution comprised in the so-called 'present cosmological paradigm': i) a stage of radiation domination from which, ii) a period of matter-radiation equality and subsequent matter domination emerge, followed by iii) a late-time inflationary stage.

\subsection{Interacting Ghost Dark Energy}\label{i-gde}

In the case when additional non-gravitational interaction between the GDE and the CDM is considered, the energy densities of GDE and CDM no longer satisfy independent conservation laws. They obey, instead

\be \dot\rho_m+3H\rho_m=Q,\;\dot\rho_{gde}+3H\rho_{gde}(1+\omega_{gde})=-Q.\label{i_cons_law}\ee As before, for convenience, the interaction term can be written as $Q=H\delta$. In the present case equation (\ref{useful}) has to be replaced by the following:

\be \frac{\dot\rho_{gde}}{H\rho_{gde}}=\frac{\dot H}{H^2}=-3(1+\omega_{gde})-\frac{\delta}{\rho_{gde}}.\label{relation'}\ee In consequence one obtains

\be 3(1+\omega_{gde})=\frac{\Omega_{gde}(3-3\Omega_{gde}+\Omega_r)-2\Omega_\delta}{\Omega_{gde}(2-\Omega_{gde})}.\label{useful'}\ee Hence,

\bea &&\omega_{gde}=-\frac{\Omega_{gde}(3-\Omega_r)+2\Omega_\delta}{3\Omega_{gde}(2-\Omega_{gde})},\notag\\
&&q=\frac{1-2\Omega_{gde}+\Omega_r-\Omega_\delta}{2-\Omega_{gde}},\notag\\
&&\frac{\dot H}{H^2}=\frac{-3+3\Omega_{gde}-\Omega_r+\Omega_\delta}{2-\Omega_{gde}}.\label{int_rel_parameters}\eea

Here, as in the former section, we will assume different forms of the interaction which are found in the bibliography \cite{iterms1,iterms2,iterms-also}. We use same variables as in \ref{n-i-gde}.

\begin{figure}[t]
\includegraphics[width=4cm,height=4cm]{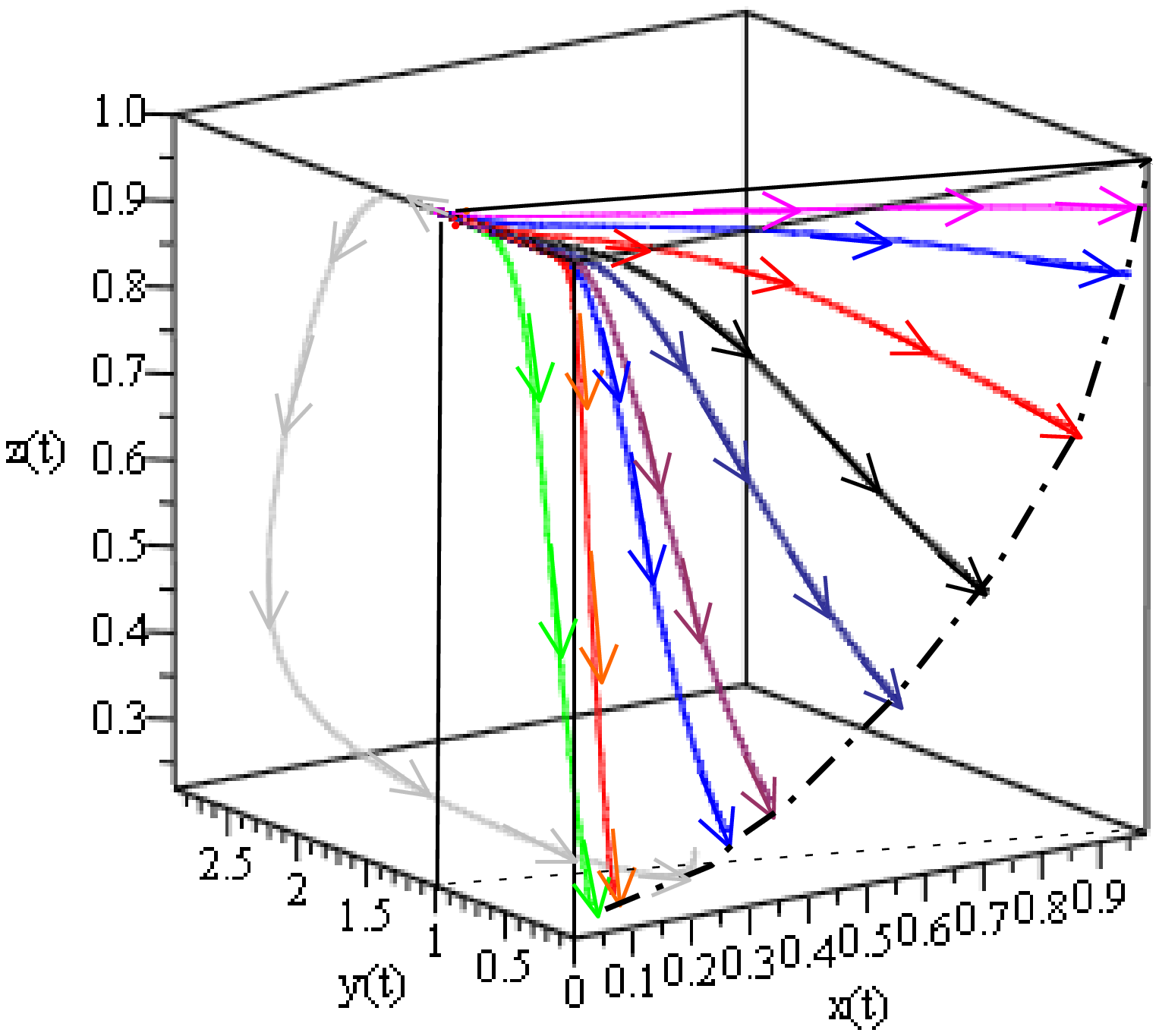}
\includegraphics[width=4cm,height=4cm]{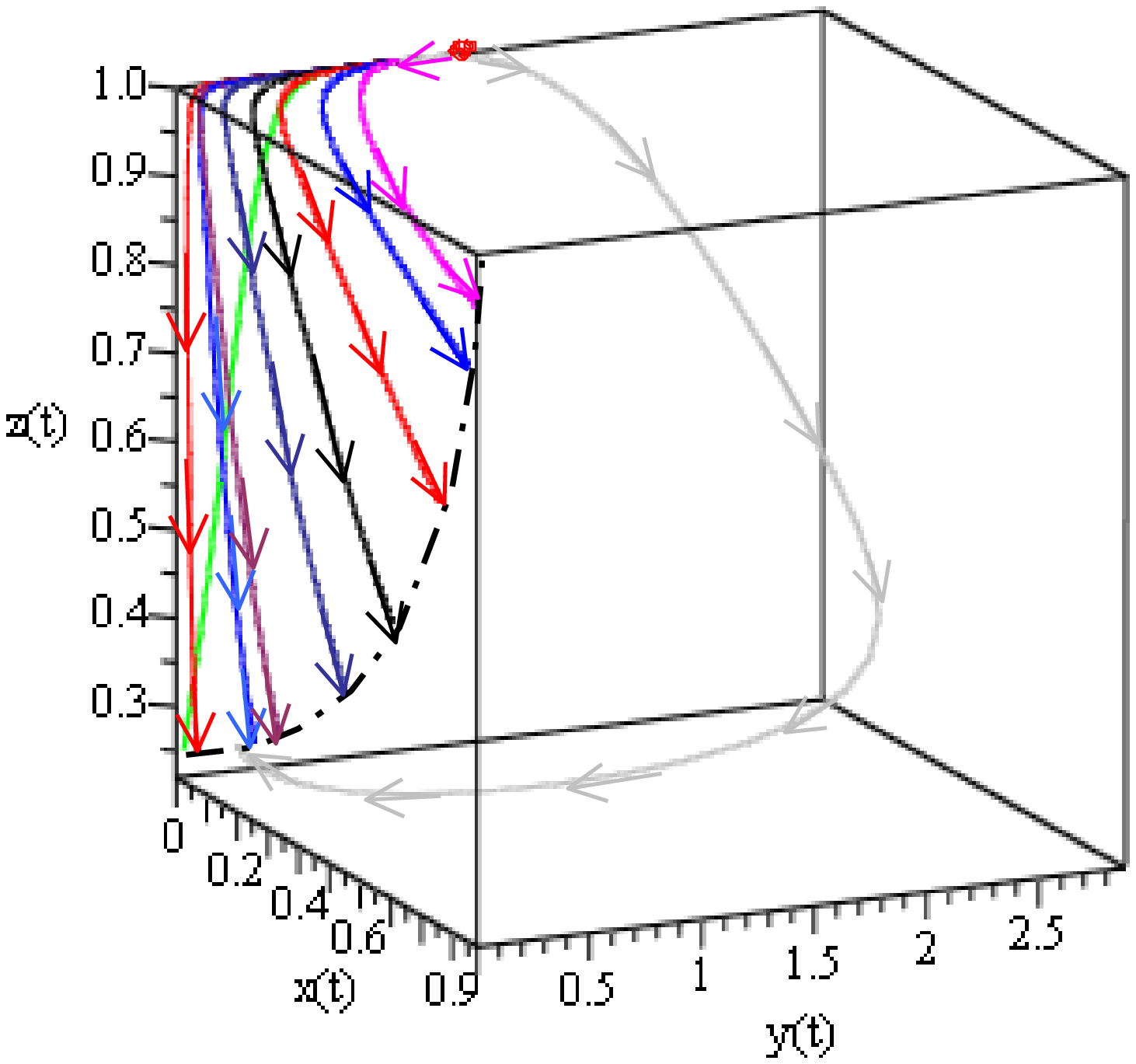}
\caption{3D ''phase portrait'' generated by given set of initial data for the autonomous system of ODE (\protect\ref{ode_xyz}). In the right-hand panel a view rotated $90^0$ clock-wise about the $z$ direction is shown. As shown in the left-hand panel, the phase space is bounded by the orthogonal planes $x=0$, $y=0$, $z=0$, $z=1$, and by the oblique plane $x+y=1$. It is seen the way the orbits of the ODE (\ref{ode_xyz}) depart from the global past attractor, point $P_r:(0,1,1)$-- radiation-dominated solution, are scattered from the saddle critical point $P_m:(0,0,1)$-- matter-dominated solution, and, depending on the initial data, these orbits approach to one of the points in the 1-dimensional manifold $M_s=\left\{\left(x,0,1/(4-3x)\right):\;0\leq x\leq 1\right\}$ depicted by the dash-doted curve joining the critical points, $P_{dS}:(1,0,1)$ and $P'_m:(0,0,1/4)$. The unphysical orbit drawn in gray color is originated by initial data leading to $z$-s smaller than $1/4$ for a finite $\tau$-interval.}
\label{fig1}
\end{figure}

\subsubsection{$\delta=\delta_0=const.$}\label{1}

Let us assume positive $\delta_0>0$ (energy transfer is from the GDE into the CDM). It is then appropriate to add a third variable (Eq.(\ref{z})) $$z=\frac{3H^2}{3H^2+\delta_0},\;0\leq z\leq 1,$$ to the variables $x$, $y$ in Eq.(\ref{xy}). The variable $z$ takes decreasing values in the above interval. Since $z=1$ corresponds to the formal limit $\delta_0\rightarrow 0$, points of the phase space lying on the plane $z=1$ are associated with non-interacting GDE solutions, unless the given fixed point can be associated with the initial big-bang singularity characterized by unbounded $H^2\rightarrow\infty$. The autonomous system of ODE obtained can be written as:

\bea &&x'=x\left[\frac{z(4-3x+y)-1}{z(2-x)}\right],\nonumber\\
&&y'=-2y\left[\frac{z(x-y)+1}{z(2-x)}\right],\nonumber\\
&&z'=-2(1-z)\left[\frac{z(4-3x+y)-1}{2-x}\right].\label{ode_xyz}\eea The phase space in this case is the following 3D finite region (see Fig.\ref{fig1}):

\bea &&\hat\Psi_{3D}=\{(x,y,z)|0\leq x\leq 1,\;0\leq y\leq 1,\nonumber\\
&&\;\;\;\;\;\;\;\;\;\;\;\;\;\;\;\;\;\;\;\;\;\;\;\;\;0\leq x+y\leq 1,\;0<z\leq 1\},\label{3d-ps}\eea where points lying on the plane $z=0$, are not included since the ODE-s (\ref{ode_xyz}) blow-up at $z=0$. Here, as before, $\Omega_m=1-x-y$, while

\bea &&\omega_{gde}=-\frac{x(3-y)z+2(1-z)}{3x(2-x)z},\nonumber\\
&&q=\frac{(y-2x)z+2z-1}{z(2-x)}.\label{rel_xyz}\eea

The equilibrium points, $P_{c_i}:(x_{c_i},y_{c_i},z_{c_i})$, of the autonomous system of ODE (\ref{ode_xyz}), together with their relevant properties, are listed below.

\begin{enumerate}

\item (Non-interacting) Radiation-dominated solution, $P_r:(0,1,1)$. The cosmological parameters are: $\Omega_r=1$, $q=1$ ($\omega_{gde}$ is undefined). This is the past attractor in $\hat\Psi_{3D}$ since the eigenvalues of the linearization matrix are all positive: $\lambda_1=1$, $\lambda_2=2$, and $\lambda_3=4$.

\item (Non-interacting) CDM domination, $P_m:(0,0,1)$. In this case $\Omega_m=1$, $q=1/2$ ($\omega_{gde}$ is undefined as well). As long as the eigenvalues of the linearization matrix for this critical point are: $\lambda_1=-1$, $\lambda_2=3/2$, and $\lambda_3=3$, the solution is a saddle equilibrium point in $\hat\Psi_{3D}$.

\item 1-dimensional manifold, $$M_s=\left\{\left(x,0,1/(4-3x)\right):\;0\leq x\leq 1\right\},$$ whose points can be associated with interacting CDM/GDE-scaling solutions whenever, $0<x<1$, since the CDM and the GDE scale as, $\Omega_m/\Omega_{gde}=(1-x)/x$. The point $(1,0,1)\Rightarrow\Omega_{gde}=1$, which also belongs in the manifold, can be associated, instead, with a non-interacting GDE-dominated phase, while the equilibrium point, $(0,0,1/4)\Rightarrow\Omega_m=1$, corresponds to an another phase of matter domination (see below). 

For points in the manifold $M_s$, the deceleration parameter $q=-1$ (accelerated expansion), while the GDE EOS parameter is given by, $\omega_{gde}=-1/x$, so that, as $x$ decreases from $1$ to $0$, $\omega_{gde}$ unboundedly decreases from $-1$ to $-\infty$.\footnote{Recall that, with the course of expansion, since $H^2$ decreases from infinitely large values in the past, into vanishingly small values in the future, in the manifold $M_s$, $z=3H^2/(3H^2+\delta_0)$ decreases from $1$ to $1/4$, so that $x=(4z-1)/3z$ takes decreasing values in the interval $[1,0]$.} Hence, this manifold can be associated with phantom-like solutions. Equilibrium points in $M_s$ are non-hyperbolic critical points since one of the eigenvalues of the linearization (also Jacobian) matrix $$\begin{pmatrix} -\frac{3x}{2-x} & 0 & \frac{18(1-x)}{(4-3x)^2(2-x)} \\ \frac{x}{2-x} & -4 & \frac{6(1-x)}{(4-3x)^2(2-x)} \\ \frac{x(4-3x)^2}{2-x} & 0 & -\frac{6(1-x)}{2-x} \end{pmatrix},$$ is vanishing: $\lambda_1=0$, $\lambda_2=-4$, and $\lambda_3=-3$. Nevertheless, the manifold itself is stable in the sense that every orbit in the phase space end ups in $M_s$ no matter which initial conditions were given (see the figure \ref{fig1}). This manifold joints the the non-interacting GDE-dominated solution ($\Omega_{gde}=1$) $P_{dS}:(1,0,1)$, with the interacting CDM domination critical point ($\Omega_m=1$) $P'_m:(0,0,1/4)$. As $P'_m$ is approached, the GDE state parameter blows-up, $$\lim_{(x,y,z)\rightarrow (0,0,1/4)}\omega_{gde}\rightarrow -\infty,$$ so that this phase of the cosmic evolution can be associated with extreme phantom behavior.

\end{enumerate}

The phase portrait for this case ($\delta=\delta_0$) is depicted by the figure \ref{fig1}. As seen, phantom-like behavior is the generic fate of the cosmic evolution, while a de Sitter universe corresponds to only one (boundary) point in the manifold $M_s$: the non-interacting GDE-dominated solution $P_{dS}$. Although the introduction of an interaction term of the form $Q=H\delta_0$ does not modify the stability properties of the fixed points as compared with the non-interacting case (\ref{n-i-gde}), nevertheless, the phase space topology is indeed modified since the isolated fixed point corresponding to GDE domination in the non-interacting case is replaced by the stable manifold $M_s$. Points in this manifold can be associated with phantom-like solutions of the cosmological equations without the need for phantom-like matter. As a consequence, for this particular form of the interaction term $Q$, since the phase space dynamics is richer, the additional non-gravitational interaction makes the model more attractive. 

In the following cases (subsections \ref{2}, and \ref{3}), the phase space coincides with the 2D phase plane corresponding to the triangular region defined in equation (\ref{phasespace}).

\subsubsection{$\delta=3b^2\rho_m$.}\label{2}

As before $b^2$ is a positive constant. Here $\Omega_\delta=3b^2(1-x-y)$. The autonomous system of ODE obtained for this particular case is given as follows:

\bea &&x'=x\left[\frac{3(1-b^2)(1-x)+(1+3b^2)y}{2-x}\right],\nonumber\\
&&y'=-2y\left[\frac{(1+3b^2)(1-y)+(1-3b^2)x}{2-x}\right].\label{ode_2}\eea For the GDE EOS parameter $\omega_{gde}$ and the deceleration parameter $q=-(1+\dot H/H^2)$, one obtains

\bea &&\omega_{gde}=-\frac{x(3-6b^2-y)+6b^2(1-y)}{3x(2-x)},\nonumber\\
&&q=\frac{1-3b^2-(2-3b^2)x+(1+3b^2)y}{2-x}.\label{rel_par_2}\eea

The critical points for this case are:

\begin{enumerate}

\item Radiation-dominated solution $$P_r:(0,1),\;\Omega_r=1,\;q=1.$$ It is the past attractor in $\hat\Psi_{2D}$ ($\lambda_1=2$, $\lambda_2=1+3b^2$). The expansion is always decelerated in this regime. The GDE-EOS parameter $\omega_{gde}$ is undefined.

\item CDM-dominated solution $$P_m:(0,0),\;\Omega_m=1,\;q=(1-3b^2)/2.$$ Here $\omega_{gde}$ is also undefined. This phase of the cosmic expansion is accelerated whenever, $b^2>1/3$. For this case the eigenvalues of the Jacobian matrix are, $\lambda_1=3(1-b^2)/2$, and, $\lambda_2=-1-3b^2$, so that, it is a saddle equilibrium point if, $b^2<1$. Otherwise, for $b^2>1$, it is the future attractor in the phase space, $\hat\Psi_{2D}$.

\item GDE-dominated (de Sitter) solution $$P_{dS}:(1,0),\;\Omega_{gde}=1,\;\omega_{gde}=-1,\;q=-1.$$ Since the eigenvalues of the Jacobian matrix for this point are $\lambda_1=-3(1-b^2)$ and $\lambda_2=-4$, it is the future attractor whenever $b^2<1$, while for $b^2>1$ it is a saddle critical point instead. 

\end{enumerate}

Addition of the non-gravitational interaction, $Q=3b^2\rho_m$, between CDM and GDE, does not modify the topology of the phase space in any significant way. The only effect of the additional interaction is to slightly modify the stability properties of the already existing fixed points in the non-interacting case. However, a very interesting thing happened: the CDM-dominated phase can be associated with accelerated expansion if $b^2>1/3$. Besides, if $b^2>1$, the stage of matter dominance not only supports accelerated expansion, but it is the future attractor in the phase space. In this case the GDE-dominated solution represents a transient stage of the cosmic evolution.\footnote{It would be interesting to show that the GDE-dominated critical point can be associated with early-time inflation for $b^2>1$.}

\subsubsection{$\delta=3b^2(\rho_m+\rho_{gde})$.}\label{3}

In this case $\Omega_\delta=3b^2(1-y)$. We obtain the following autonomous system of ODE:

\bea && x'=x\left[\frac{3(1-b^2)+(3b^2+1)y-3x}{2-x}\right],\notag\\
&& y'=-2y\left[\frac{(3b^2+1)(1-y)+x}{2-x}\right].\label{ode_3}\eea 

As in the former case, here the phase space is the same as in Eq.(\ref{phasespace}). The magnitudes of relevance can be written in terms of the phase space variables ($\Omega_m=1-x-y$):

\bea &&\omega_{gde}=-\frac{x(3-y)+6b^2(1-y)}{3x(2-x)},\nonumber\\
&&q=\frac{1-3b^2-2x+(1+3b^2)y}{2-x}.\label{rel_int}\eea

The following equilibrium points are found:

\begin{enumerate}

\item Non-interacting, radiation-dominated phase $$P_r:(0,1),\;\Omega_r=1,\;\Omega_\delta=0,$$ while $\omega_{gde}$ is undefined. Since the eigenvalues of the linearization matrix, $\lambda_1=2$, $\lambda_2=1+3b^2$, are both positive, this is a unstable critical point (a past attractor), which corresponds to a decelerating expansion solution ($q=1$). 
  
\item CDM-dominated phase $$P_m:(0,0),\;\Omega_m=1,\;q=(1-3b^2)/2.$$ As before the GDE-EOS parameter $\omega_{gde}$ is undefined in this case. The eigenvalues of the linearization matrix corresponding to this critical point are: $\lambda_1=-1-3b^2$, $\lambda_2=-3/2(b^2-1)$, so that this solution is a saddle critical point if $b^2<1$, while, for $b^2>1$, it is the future attractor in, $\hat\Psi_{2D}$. For $b^2>1/3$ the solution is inflationary ($q<0$).
  
\item CDM/GDE-scaling solution $P_{m-sc}:(1-b^2,0)$. This equilibrium point exists whenever $b^2<1$ (for $b^2=1$ this point coincides with $P_m$ above). In this case the energy densities of DM and of GDE scale as $\Omega_m/\Omega_{gde}=b^2/(1-b^2)$. It is a late-time inflationary solution ($q=-1$). The eigenvalues of the linearization matrix for $P_{m-sc}$ are: $\lambda_1=-4$, $\lambda_2=3-6/(1+b^2)$, so that, since $b^2<1$, this solution is the late-time (future) attractor. The GDE-EOS parameter $\omega_{gde}=1/(b^2-1)\;\Rightarrow\;-\infty<\omega_{gde}<-1$. I. e., in this regime the GDE always behaves like phantom matter. Since, according to the observational evidence, at present $$0\leq\frac{\Omega_m}{\Omega_{gde}}\leq 1\;\Rightarrow\;\frac{b^2}{1-b^2}\leq 1,$$ then $b^2<1/2$.
  
\end{enumerate}

Consideration of an interaction term of the form $Q=3b^2H(\rho_m+\rho_{gde})$ does actually modify the topology of the phase space (\ref{phasespace}) by replacing the GDE-dominated (de Sitter) solution in the non-interacting case by a CDM/GDE scaling solution. Hence, the additional non-gravitational interaction between the cold dark matter and the ghost dark energy, in this case, smooths out the coincidence problem. Actually, since the scaling solution, whenever it exists, is the (global) future attractor, then, independent on the initial conditions chosen, scaling between dark matter and dark energy always dictates the late-time cosmological behavior. Besides, at the CDM/GDE scaling point the expansion is accelerated since $q=-1$.

\section{GDE plus a decaying $\Lambda$}

Here we consider a further generalization of the GDE model explored so far (Eq.(\ref{sim-h})), which is given by \cite{chinos'} (see also \cite{also}) 

\be \rho_{gde}=\alpha H+\beta H^2,\label{sim-h+h2}\ee where, in order to warrant positivity of energy, the constant parameter $\beta\geq 0$ should be a non-negative quantity. The second term in the above expression is inspired in models of decaying cosmological constant \cite{decaying-lambda}. One may expect that the subleading term $H^2$ in (\ref{sim-h+h2}) might play a crucial role in the early evolution of the universe \cite{maggiore}. It was shown that considering this term can give better agreement with observational data compared to the usual GDE \cite{chinos'}.

\subsection{Non-interacting model}\label{ni-gde+l}

The cosmological field equations are the same as in \ref{n-i-gde} (equations (\ref{friedman_eq}), (\ref{conservation_law})). Here we use same variables $x$ and $y$ defined in (\ref{xy}). Notice that the relationship (\ref{sim-h+h2}) can always be written in the following alternative form:

\be \frac{\alpha}{H}=3x-\beta.\label{util}\ee It is seen from (\ref{util}) that, since, $$x=\frac{1}{3}\left(\beta+\frac{\alpha}{H}\right),\;0\leq x\leq 1,$$ then $H$ is bounded from below: $$0\leq\beta+\frac{\alpha}{H}\leq 3\;\Rightarrow\;H\geq H_0=\frac{\alpha}{3-\beta}.$$ This entails that the de Sitter fixed point can describe the late-time cosmological dynamics in this model, while the decaying cosmological constant can drive the early-times dynamics since, near of the initial singularity $$H\rightarrow\infty\;\Rightarrow\;x\rightarrow\frac{\beta}{3},\Rightarrow\;\alpha\rightarrow 0\;\Rightarrow\;\rho_{gde}=\beta H^2.$$

Given that the free parameter $\alpha\geq 0$ is non-negative and, since we are interested only in expanding solutions $H\geq 0$, then, from (\ref{util}) it follows that $$\frac{\alpha}{H}=3x-\beta\geq 0\;\Rightarrow\;x\geq\frac{\beta}{3}.$$ In consequence, the phase space of this model will correspond to the compact triangular region $\hat\Psi^*_{2D}\subset\hat\Psi_{2D}$:

\bea &&\hat\Psi^*_{2D}=\{(x,y)|\;\beta/3\leq x\leq 1,\;0\leq y\leq 1,\nonumber\\
&&\;\;\;\;\;\;\;\;\;\;\;\;\;\;\;\;\;\;\;\;\;\;\;\;\;\;\;\;\;\;\;\;\;\;\;\;\;\;\;\;\;\;\;\;\;\;0\leq x+y\leq 1\}.\label{phasespace'}\eea 

Due to (\ref{sim-h+h2}) equation (\ref{relation}) is replaced by the following equation:

\be \frac{\dot\rho_{gde}}{H\rho_{gde}}=\left(\frac{3x+\beta}{3x}\right)\frac{\dot H}{H^2}=-3(1+\omega_{gde}).\label{rel}\ee The corresponding autonomous system of ODE is now given by,

\bea &&x'=(\beta-3x)\left[\frac{3(x-1)-y}{6-\beta-3x}\right]\nonumber\\
&&y'=-2y\left[\frac{3(1+x-y)-2\beta}{6-\beta-3x}\right].\label{xy-new}\eea As before $\Omega_m=1-x-y$, while the expressions for the GDE-EOS and the deceleration parameters, are modified

\bea &&\omega_{gde}=-1-\frac{(3x+\beta)[3(x-1)-y]}{3x(6-\beta-3x)},\nonumber\\
&&q=\frac{3+\beta-6x+3y}{6-\beta-3x}.\nonumber\eea

The following fixed points are found in $\hat\Psi^*_{2D}$:

\begin{enumerate}

\item Radiation/GDE-scaling solution $$P_{r-sc}:\left(\beta/3,1-\beta/3\right),\;q=1.$$ The GDE EOS parameter $\omega_{gde}=1/3$ whenever $\beta\neq 0$. For $\beta=0$ it is undefined instead. This point exists whenever $\beta\leq 3$. Given that the eigenvalues of the corresponding Jacobian matrix are positive: $\lambda_1=2$, $\lambda_2=1$, it is a global past attractor (source point) in $\hat\Psi^*_{2D}$. The ghost dark energy tracks radiation with their dimensionless energy density parameters scaling as $$\frac{\Omega_r}{\Omega_{gde}}=\frac{3-\beta}{\beta}.$$

\item CDM/GDE-scaling solution $$P_{m-sc}:\left(\beta/3,0\right),\;\omega_{gde}=0,\;q=1/2,$$ which exists only for $\beta\leq 3$. This case is similar to the former one: ghost dark energy tracks cold dark matter and $$\frac{\Omega_m}{\Omega_{gde}}=\frac{3-\beta}{\beta},$$ but, this time, it is a saddle critical point since the eigenvalues of the linearization matrix for this point are of opposite sign: $\lambda_1=3/2$, $\lambda_2=-1$.

\item de Sitter (GDE-dominated) phase $$P_{dS}:(1,0),\;\omega_{gde}=-1,\;q=-1.$$ This is an inflationary solution. The ghost dark energy behaves like a cosmological constant. This solution is a global (future) attractor in the phase space $\hat\Psi^*_{2D}$, since both eigenvalues of the corresponding Jacobian matrix: $\lambda_1=-3$ and $\lambda_2=-4$, are negative. This means that $P_{dS}$ is the end-point of any phase space orbit and, hence, it depicts always the late-time cosmological dynamics. This is consistent with the analysis of the expression (\ref{util}) at the beginning of this subsection, according to which the Hubble parameter $H$, is always bounded from below ($H\geq H_0$).

\end{enumerate}

Take a look at (\ref{util}), it is then clear that, for points where $x=\beta/3$ (fixed points $P_{r-sc}$ and $P_{m-sc}$), the ghost dark energy behaves like a decaying cosmological constant $\rho_{gde}=\beta H^2$, while for the GDE-dominated de Sitter solution $P_{dS}$, since $x=1$, the GDE behaves like a standard (truly constant) cosmological constant: 

\bea &&H=H_0=\frac{\alpha}{3-\beta}\;\Rightarrow\nonumber\\
&&\rho_{gde}=\alpha H_0+\beta H_0^2=\frac{3\alpha^2}{(3-\beta)^2}=\Lambda.\nonumber\eea

\subsection{Interacting Model}

We keep using the same phase space variables, $x=\Omega_{gde}$, $y=\Omega_r$. In this case the conservation equations (\ref{conservation_law}) are replaced by (\ref{i_cons_law}) (the conservation equation for radiation is unchanged). The relation (\ref{rel}), distinctive of ghost dark energy, transforms into the following relationship, 

\be \frac{\dot\rho_{gde}}{H\rho_{gde}}=\left(\frac{3x+\beta}{3x}\right)\frac{\dot H}{H^2}=-3(1+\omega_{gde})-\frac{\Omega_\delta}{x},\label{rel'}\ee where $\Omega_\delta\equiv\delta/3H^2$. Besides,

\be \frac{1}{3}\frac{\dot H}{H^2}=\frac{3(x-1)-y+\Omega_\delta}{6-\beta-3x}.\label{doth}\ee Also, 

\be \omega_{gde}=-1-\frac{(3x+\beta)[3(x-1)-y]+6\Omega_\delta}{3x(6-\beta-3x)}.\label{wgde}\ee

A very general form of the interaction term can be $$\delta=\delta_0+3b^2(\rho_m+\rho_{gde}),$$ which comprises in a single expression the interaction terms of \ref{q-1}, \ref{q-2}, and \ref{q-3}. However, as in the former sections, it is convenient to study separately the different particular cases.

\subsubsection{$\delta=\delta_0$}\label{i-gde-1}

In the present case the already defined variables $$x=\Omega_{gde},\;y=\Omega_r,\;z=\frac{3H^2}{3H^2+\delta_0},$$ are a good set of phase space variables in order to write the cosmological equations (\ref{friedman_eq}), (\ref{i_cons_law}), plus the conservation equation for the radiation component, in the form of an autonomous system of ODE. In terms of these variables the dimensionless energy density associated with the interaction $\Omega_\delta=\Omega_{\delta_0}=\delta_0/3H^2$, can be written as

\be \Omega_{\delta}=\frac{1-z}{z},\label{i-dens-1}\ee while the GDE-EOS and the deceleration parameters can be written as $$\omega_{gde}=-1-\frac{(3x+\beta)[3(x-1)-y]z+6(1-z)}{3x(6-\beta-3x)z},$$ and $$q=-1-3\left\{\frac{[3(x-1)-y]z+1-z}{(6-\beta-3x)z}\right\},$$ respectively. The autonomous system of ODE for this case is the following:

\bea &&x'=(\beta-3x)\left\{\frac{[3(x-1)-y]z+1-z}{(6-\beta-3x)z}\right\},\nonumber\\
&&y'=-2y\left[\frac{3(1+x-y)z+1-z-2\beta z}{(6-\beta-3x)z}\right],\nonumber\\
&&z'=6z(1-z)\left\{\frac{[3(x-1)-y]z+1-z}{(6-\beta-3x)z}\right\},\label{i-xyz-new}\eea while the phase coincides with the 3D compact region $\hat\Psi^*_{3D}\subset\hat\Psi_{3D}$:

\bea &&\hat\Psi^*_{3D}=\{(x,y,z)|\;\beta/3\leq x\leq 1,\;0\leq y\leq 1,\nonumber\\
&&\;\;\;\;\;\;\;\;\;\;\;\;\;\;\;\;\;\;\;\;\;\;\;\;\;\;\;0\leq x+y\leq 1,\;0<z\leq 1\}.\label{3d'-ps}\eea 

The following critical points/manifolds can be found in the phase space $\hat\Psi^*_{3D}$:

\begin{enumerate}

\item (Non-interacting) radiation/GDE-scaling point, $$P_{r-sc}:\left(\beta/3,1-\beta/3,1\right),\;q=1,$$ where the dimensionless energy densities scale as, $$\frac{\Omega_r}{\Omega_{gde}}=\frac{3-\beta}{\beta}.$$ Whenever $\beta\neq 0$, the GDE EOS parameter $\omega_{gde}=1/3$ (if $\beta=0$ it is undefined). This fixed point exists if $\beta\leq 3$. This is either a non-interacting point, or it is associated with the initial big-bang singularity instead ($H^2\rightarrow\infty$ is unbounded) since $z=1\;\Rightarrow\;(\delta_0/3H^2)\rightarrow 0$. The eigenvalues of the linearization matrix for this point are: $\lambda_1=1$, $\lambda_2=2$, and, $\lambda_3=1/3$, so that $P_{r-sc}$ is a global past attractor in $\hat\Psi^*_{3D}$.

\item (Non-interacting) CDM/GDE-scaling fixed point, $$P_{m-sc}:\left(\beta/3,0,1\right),\;\omega_{gde}=0,\;q=1/2,$$ which exists whenever, $\beta\leq 3$. The CDM and GDE dimensionless energy density parameters scale in a constant fraction, $$\frac{\Omega_m}{\Omega_{gde}}=\frac{3-\beta}{\beta},$$ and GDE tracks CDM since, $\omega_{gde}$ $=\omega_m$ $=0$. Given that, the eigenvalues of the Jacobian matrix for, $P_{m-sc}$, are of opposite sign: $\lambda_1=3$, $\lambda_2=3/2$, $\lambda_3=-1$, this is a saddle critical point in $\hat\Psi^*_{3D}$.

\item 1-dimensional compact manifold, $$M_s=\left\{\left(x,0,\frac{1}{4-3x}\right):\;\frac{\beta}{3}\leq x\leq 1\right\},$$ where $\omega_{gde}=-1/x$, $q=-1$. Besides, $$\frac{\Omega_m}{\Omega_{gde}}=\frac{1-x}{x}.$$ Since (the real part of) one of the eigenvalues of the Jacobian matrix for points in this manifold, $P_{s_i}\in M_s$, vanishes: $$\lambda_1=0,\;\lambda_2=-3,\;\lambda_3=-\frac{4(3-\beta)}{3-\beta+3(1-x)},$$ then, the $P_{s_i}$, are non-hyperbolic critical points. This means that no final judgment can be made about their stability as isolated points. We want to underline, however, that the manifold $M_s$ itself is stable in the sense that, no matter which initial conditions to choose, every orbit in the phase space will, necessarily, end-up in a given point belonging in $M_s$. 

For, $\beta/3<x<1$, fixed points, $P_{s_i}\in M_s$, correspond to (inflationary) phantom-like ($\omega_{gde}<-1$), interacting CDM/GDE-scaling solutions. 

The end-points of this manifold are: 

\begin{enumerate}

\item (Interacting) CDM/GDE-scaling fixed point, $$P^*_{m-sc}:\left(\frac{\beta}{3},0,\frac{1}{4-\beta}\right),\;\omega_{gde}=-\frac{3}{\beta},\;q=-1,$$ for which, $\Omega_m/\Omega_{gde}=(3-\beta)/\beta$. Unlike the (non-interacting) matter-scaling solution, $P_{m-sc}$, since, in this case, $q=-1$, this fixed point is associated with inflationary expansion. Besides, even if the cold dark matter and the ghost dark energy scale in a constant fraction, due to the additional non-gravitational interaction, the GDE does not track the CDM: $$\omega_{gde}=-\frac{3}{\beta}\;\neq\;\omega_m=0.$$

\item (Non-interacting) de Sitter solution, $$P_{dS}=\left(1,0,1\right),\;\omega_{gde}=-1,\;q=-1.$$ Since, $\Omega_{gde}=1$, this is a GDE-dominated solution. In this case, according to (\ref{util}), $$H=H_0=\frac{\alpha}{3-\beta}.$$

\end{enumerate}

\end{enumerate}

There exists another 1-dimensional critical manifold, $$M^*_s=\left\{\left(x,3-\beta,\frac{1}{7-\beta-3x}\right):\;\frac{\beta}{3}\leq x\leq\frac{6-\beta}{3}\right\},$$ for which, $$\Omega_m=\beta-2-x,\;\omega_{gde}=-\frac{6-\beta}{3x},\;q=-1.$$ However, for this manifold to be physically meaningful, it is required that, i) $\beta\leq 3$, and, ii) $(6-\beta)/3\leq 1$.\footnote{The latter constraint comes from requiring, $x\leq 1$.} Both constraints, together taken lead to, $\beta=3$. Therefore, the only physically meaningful fixed point in $M^*_s$ is the de Sitter solution, $P_{dS}$, already included in the manifold $M_s$.

\subsubsection{$\delta=3b^2\rho_m$}\label{i-gde-2}

The phase space for this and the following cases of interacting CDM and GDE is the two-dimensional triangular region, $\hat\Psi^*_{2D}$, defined in Eq.(\ref{phasespace'}). Since, in the present case, $\delta=3b^2\rho_m$, we have,

\be \Omega_\delta=3b^2(1-x-y),\label{i-dens-2}\ee so that,

\bea &&\omega_{gde}=-1-\frac{(3x+\beta)[3(x-1)-y]}{3x(6-\beta-3x)}\nonumber\\
&&\;\;\;\;\;\;\;\;\;\;\;\;\;\;\;\;\;\;\;\;\;\;\;\;\;\;\;\;\;\;\;\;\;\;\;\;\;\;\;\;\;-\frac{6b^2(1-x-y)}{x(6-\beta-3x)},\nonumber\\
&&q=-1-3\left[\frac{3(x-1)-y+3b^2(1-x-y)}{6-\beta-3x}\right].\nonumber\eea

We are led to the following autonomous system of ODE:

\bea &&x'=(\beta-3x)\left[\frac{3(x-1)-y+3b^2(1-x-y)}{6-\beta-3x}\right],\nonumber\\
&&y'=-6y\left[\frac{1+x-y+3b^2(1-x-y)-\frac{2\beta}{3}}{6-\beta-3x}\right].\label{i-xy-new}\eea

This time, basically the same three fixed points of subsection \ref{ni-gde+l}, can be found in the phase space $\hat\Psi^*_{2D}$, however, their stability properties, among others, are slightly modified by the interaction:

\begin{enumerate}

\item Radiation/GDE-scaling solution, $$P_{r-sc}:\left(\beta/3,1-\beta/3\right),\;q=1,$$ where $$\frac{\Omega_r}{\Omega_{gde}}=\frac{3-\beta}{\beta}.$$ For $\beta\neq 0$, $\omega_{gde}=1/3$. This fixed point exists if $\beta\leq 3$. The eigenvalues of the Jacobian matrix are both positive: $\lambda_1=2$, and, $\lambda_2=1+3b^2$, so that, as in the non-interacting case, the point, $P_{r-sc}$, is a past attractor in the phase space.

\item CDM/GDE-scaling phase, $$P_{m-sc}:\left(\beta/3,0\right),\;\omega_{gde}=-3b^2/\beta,\;q=(1-3b^2)/2,$$ which, as with the former critical point, exists whenever $\beta\leq 3$. As in the non-interacting case, the energy densities involved scale in a constant fraction: $$\frac{\Omega_m}{\Omega_{gde}}=\frac{3-\beta}{\beta}.$$ However, this time the ghost dark energy does not track the cold dark matter, since, in general ($b^2\neq 0$, $\beta\neq 0$), $\omega_{gde}\neq\omega_m=0$. Besides, for $b^2>1/3$, this solution depicts accelerated expansion, while, for $b^2>\beta/3$, the ghost dark energy behaves like a phantom field (without the unwanted features of a phantom field, of course). 

The stability is also modified by the interaction in respect to the non-interacting case, since the eigenvalues of the corresponding linearization matrix: $\lambda_1=3(1-b^2)/2$, $\lambda_2=-3(1+b^2)$. Hence, whenever, $b^2>1$, the CDM/GDE-scaling (inflationary) solution is the future attractor in the phase space.

\item de Sitter solution, $$P_{dS}:(1,0),\;\Omega_{gde}=1,\;\omega_{gde}=-1,\;q=-1.$$ In this case the additional non-gravitational interaction, $\delta=3b^2\rho_m$, affects only the stability properties, since, $\lambda_1=-4$, and $\lambda_2=-3(1-b^2)$. It is seen that, for $b^2>1$, the de Sitter solution is a saddle critical point in $\hat\Psi^*_{2D}$, while, for $b^2<1$, it is the future attractor instead.

\end{enumerate}

\subsubsection{$\delta=3b^2(\rho_m+\rho_{gde})$}\label{i-gde-3}

As before we use variables $x=\Omega_{gde}$, and $y=\Omega_r$, taking values in the compact triangular region, $\hat\Psi^*_{2D}$ (\ref{phasespace'}). In this case the dimensionless 'energy density' of the interaction, $\Omega_\delta=3b^2(1-y)$. For the GDE-EOS and deceleration parameters we get:

\bea &&\omega_{gde}=-1-\frac{(3x+\beta)[3(x-1)-y]}{3x(6-\beta-3x)}\nonumber\\
&&\;\;\;\;\;\;\;\;\;\;\;\;\;\;\;\;\;\;\;\;\;\;\;\;\;\;\;\;\;\;\;\;\;\;\;-\frac{6b^2(1-y)}{x(6-\beta-3x)},\nonumber\\
&&q=-1-3\left[\frac{3(x-1)-y+3b^2(1-y)}{6-\beta-3x}\right].\nonumber\eea

We obtain the following autonomous system of ODE:

\bea &&x'=(\beta-3x)\left[\frac{3(x-1)-y+3b^2(1-y)}{6-\beta-3x}\right],\nonumber\\
&&y'=-6y\left[\frac{1+x-y+3b^2(1-y)-\frac{2\beta}{3}}{6-\beta-3x}\right].\label{i-xy'-new}\eea

The non-gravitational interaction, in this case, appreciably modifies the topology of the phase space with respect to the non-interacting case. The equilibrium points found in, $\hat\Psi^*_{2D}$, are the following:

\begin{enumerate}

\item CDM/GDE-scaling solution, $P_{m-sc}:\left(\beta/3,0\right)$, $$\omega_{gde}=-\frac{9b^2}{\beta(3-\beta)},\;q=\frac{3-\beta-9b^2}{2(3-\beta)}.$$ This fixed point exists whenever $\beta\leq 3$. The eigenvalues of the linearization matrix are: $$\lambda_1=-1-\frac{9b^2}{3-\beta},\;\lambda_2=\frac{3}{2}\left(1-\frac{3b^2}{3-\beta}\right),$$ so that $P_{m-sc}$ is stable if $3(1-b^2)<\beta<3$. Otherwise it is a saddle.

\item Interacting CDM/GDE-scaling solution, $$P^*_{m-sc}:(1-b^2,0),\;\omega_{gde}=-\frac{1}{1-b^2},\;q=-1,$$ which exists if $b^2\leq 1$. The eigenvalues of the linearization matrix are: $$\lambda_1=-4,\;\lambda_2=-\frac{3(3-\beta-3b^2)}{3-\beta+3b^2}.$$ This critical point is stable whenever, either $\beta<3(1-b^2)$, or $\beta>3(1+b^2)$. If $3(1-b^2)<\beta<3(1+b^2)$ it is a saddle point.

\end{enumerate}

There is another critical point, $$P_{r-m-sc}:\left(\frac{\beta}{3},\frac{3-\beta+9b^2}{3(1+3b^2)}\right),$$ but, in this case, $\Omega_m=-\frac{\beta b^2}{1+3b^2}<0$, which is not physical, so this point does not belong in $\hat\Psi^*_{2D}$.

\section{Discussion}

Consideration of non-minimal coupling between the matter degrees of freedom and the scalar field, or, alternatively, of additional non-gravitational interactions between matter and the dark energy, has been repeatedly invoked to get late-time matter-scaling attractors which allow accelerated expansion \cite{amendola,interaction,nm-coupling}. A scalar field is expected to couple explicitly to ordinary matter unless some special symmetry prevents or suppresses the coupling \cite{amendola,carroll}. For the case studied in section \ref{quintessence} of this paper, i. e., scalar (quintessence) field with an exponential potential, even in the minimal coupling situation (studied in Ref.\cite{wands}), there exists a matter-scaling late-time attractor, however it is a decelerated solution in contradiction with the present cosmological paradigm. As we have shown here, consideration of the non-gravitational interaction term $Q$ may lead to CDM/quintessence scaling critical point which is correlated with accelerated expansion if, either $Q=3b^2H\rho_m$, or $Q=3b^2H(\rho_m+\rho_\vphi)$. For the case when $Q=H\delta_0$ the matter-scaling solution is associated with decelerated expansion as it happens in the non-interacting case. 

At the same time the non-minimal coupling of CDM to the quintessence field leads to unwanted effects. Depending on the kind of interaction considered the modifications to the dynamics might be either: i) minimal if $Q=H\delta_0$ since only the stability is modified but not the asymptotic structure of the phase space, or ii) substantial if $Q=3b^2H\rho_m$ or $Q=3b^2H(\rho_m+\rho_\vphi)$: the asymptotic structure is modified by the interaction since one or several critical points are erased. In the two latter cases the CDM-dominated point is erased, while the CDM/quintessence scaling solution is transformed from stable into a saddle critical point in the phase space, i. e., matter-scaling results in a transient stage of the cosmic evolution. Besides, at the matter-scaling point, even if the energy densities of the dark matter and of the ghost dark energy are related in a constant fraction $\Omega_{cdm}/\Omega_{gde}=const$, the quintessence field does not track the CDM equation of state parameter: $\omega_\vphi\neq\omega_m=0$. This means that the resulting model may not take account for the observed amount of cosmic structure. The following conclusion may be read off from the former result: since the additional non-gravitational interaction $Q$ modifies the phase space topology in the previously discussed way, a suitable model of dark energy is transformed by the interaction into a model that may be ruled out by the observations on structure formation.

\subsection{Ghost Dark Energy}

The situation is a bit different if the quintessence field is replaced by a QCD ghost. In this case, in general, the interaction makes the GDE model more attractive and better suited to accommodate the present cosmological paradigm without the coincidence problem. In the simplest situation when the additional non-gravitational interaction is not considered, there are 3 equilibrium points in the equivalent phase space of the GDE model. For the model $\rho_{gde}=\alpha H$ these are: i) the radiation-dominated point $P_r:(0,1)$ -- the past attractor, ii) CDM-dominated critical point $P_m:(0,0)$ -- saddle equilibrium point, and iii) de Sitter (GDE-dominated) phase $P_{dS}:(1,0)$ -- the future attractor. Meanwhile, for the model $\rho_{gde}=\alpha H+\beta H^2$ the critical points are: i) the radiation-scaling point $P_{r-sc}:(\beta/3,1-\beta/3)$, exists if $\beta\leq 3$, the GDE mimics radiation $\omega_{gde}=1/3$, $\Omega_r/\Omega_{gde}=(3-\beta)/\beta$ -- the past attractor, ii) CDM-scaling point $P_{m-sc}:(\beta/3,0)$, also exists if $\beta\leq 3$, the ghost field tracks cold dark matter $\omega_{gde}=\omega_{cdm}=0$, $\Omega_m/\Omega_{gde}=(3-\beta)/\beta$ -- a saddle point, and iii) GDE-dominated (de Sitter) point $P_{dS}:(1,0)$ -- the future attractor. As seen, the role of the term $\propto H^2$ is to replace radiation domination and matter domination by radiation scaling and matter scaling respectively. This is expected since the QCD ghost, in this case, is not vanishing at any stage of the cosmic evolution: at early stages the term $\propto\beta H^2$ dominates, while at late times it is $\alpha H$ which dominates the cosmic dynamics. The topology of the phase space is also modified by the quadratic term since there is bifurcation at the critical value $\beta=3$: for $\beta\leq 3$ the 3 critical points can be found, but for $\beta>3$ only the de Sitter point survives.

Let us discuss now what the role of the additional non-gravitational interaction is. The constant interaction $\delta=\delta_0$ modifies mainly the GDE-dominated de Sitter point in both cases, $\delta=3b^2\rho_m$ and $\delta=3b^2(\rho_m+\rho_{gde})$, by replacing it by a one-dimensional manifold $M_s$, which is characterized by scaling of CDM and GDE: $\Omega_m/\Omega_{gde}=(1-x)/x$. This manifold is correlated with accelerated expansion ($q=-1$) and joints a matter-dominated (accelerated) phase ($x=0$), with a GDE dominated one ($x=1$). Since $\omega_{gde}=-1/x$, points in the critical manifold are associated with phantom QCD GDE. Although critical points in $M_s$ represent scaling of the dark matter and of the ghost dark energy, which represent accelerated expansion, the coincidence problem is not alleviated by the interaction $Q=H\delta_0$. Actually, even if every phase space orbit, no matter what the initial conditions are, ends up at a point in $M_s$, the given constant ratio $\Omega_m/\Omega_{gde}$ is highly dependent on the initial conditions.

If consider the GDE model with $\rho_{gde}=\alpha H$, an interaction term of the form $Q=3b^2H\rho_m$ modifies the properties of the matter dominance point $P_m:(0,1)$, $\Omega_m=1$, which is associated with decelerated expansion ($q=1/2$) in the uncoupled case, not only by allowing a range in parameter space $b^2>1$ where matter dominance is the late-time attractor, but also by allowing a range $b^2>1/3$ where accelerated expansion may be associated to dark matter domination. For $b^2>1$ the GDE-dominated de Sitter solution $P_{dS}:(1,0)$ ($\omega_{gde}=-1$, $q=-1$) is a saddle critical point in the phase space (it is the future attractor otherwise). In the case when the interaction term $Q=3b^2H(\rho_m+\rho_{gde})$ is assumed, the modified matter-dominated critical point shares the same properties with the case of interaction $\delta=3b^2\rho_m$ (future attractor for $b^2>1$, associated with accelerated expansion if $b^2>1/3$). In addition the de Sitter phase in the uncoupled case is replaced by matter-scaling $P_{dS}\rightarrow P_{m-sc}:(1-b^2,0)$, $\Omega_m/\Omega_{gde}=b^2/(1-b^2)$, which is associated with accelerated expansion ($q=-1$), and exists if $b^2<1$. It is the late-time attractor whenever it exists. Hence, in this case, as long as $b^2<1$, the coincidence problem is actually resolved by the additional non-gravitational interaction.

For the QCD GDE model with $\rho_{gde}=\alpha H+\beta H^2$, in the case of interaction $Q=3b^2\rho_m$, the stability of the matter-scaling point $P_{m-sc}$, $\Omega_m/\Omega_{gde}=(3-\beta)/\beta$, is modified for $b^2>1$, in which case $P_{m-sc}$ can be the late-time attractor. Besides, for $b^2>1/3$ this point is correlated with accelerated expansion and a resolution of the coincidence problem is also achieved. Notice that, since $\omega_{gde}=-3b^2/\beta\neq 0$, the ghost dark energy does not mimic dark matter ($\omega_{cdm}=0$). It is also noticeable also that the matter-scaling point exists whenever $\beta<3$. The situation is not as attractive if consider the interaction term $Q=3b^2(\rho_m+\rho_{gde})$. In this case only CDM-scaling solutions survive: the radiation-scaling point is erased, while the de Sitter point is replaced by a new phase of matter scaling $P_{dS}\rightarrow P^*_{m-sc}:(1-b^2,0)$. We are left with two different phases of dark matter/ghost dark energy scaling behavior: i) $P_{m-sc}:(\beta/3,0)$ which exists if $\beta\leq 3$ and is the future attractor if $3(1-b^2)<\beta<3$, and ii) $P^*_{m-sc}:(1-b^2,0)$ which exists if $b^2\leq 1$ and is the late-time attractor if either $\beta<3(1-b^2)$, or $\beta>3(1+b^2)$. $P_{m-sc}$ is correlated with accelerated expansion whenever $3(1-3b^2)<\beta<3$, while $P^*_{m-sc}$ depicts always accelerated expansion since $q=-1$. Even if the coincidence is resolved in this case by the non-minimal coupling, the resulting model might be ruled out by the observational evidence on structure formation since there are not critical points that can be correlated neither with radiation nor with dark matter dominance.

\section{Conclusions}

We have performed a detailed study of the phase space of the QCD ghost dark energy models where $\rho_{gde}=\alpha H$ and $\rho_{gde}=\alpha H+\beta H^2$ respectively. Both cases with and without additional non-gravitational interaction between cold dark matter and the Veneziano ghost field have been scrutinized. Special attention has been paid to the correct choice of the phase space variables which allow for bounded and compact spaces in each case. As a result all of the possible critical points of physical relevance were found. It has been shown that the non-minimal coupling introduces more feasible changes in the asymptotic structure of the equivalent phase space of QCD GDE models than for quintessence models with an exponential potential. 

In this paper we have neglected the contribution from the dynamics of the trapping horizon as it is done in most works on GDE \cite{urban,ariel,ohta,chinos,also,instability,alberto-rozas,other,gde-thermodynamics,iterms-also}. This particular case, strictly speaking, corresponds to dark energy dominance so that, in general, the latter contribution to the dynamics has to be taken under consideration. We defer investigation of this more realistic case for further work.

Useful remarks by S Odintsov and S Nojiri are acknowledged. The authors thank SNI of Mexico for support. I Q also thanks the mathematics department at CUCEI, Universidad de Guadalajara, for partial support.

\end{document}